\documentclass[fleqn,usenatbib,useAMS]{mnras}
\usepackage[dvipsnames]{xcolor}
\usepackage{tikz,hyperref}

\definecolor{lime}{HTML}{A6CE39}
\DeclareRobustCommand{\orcidicon}{
	\begin{tikzpicture}
	\draw[lime, fill=lime] (0,0) 
	circle [radius=0.13] 
	node[white] {{\fontfamily{qag}\selectfont \tiny ID}};
	\draw[white, fill=white] (-0.0625,0.095) 
	circle [radius=0.007];
	\end{tikzpicture}
	\hspace{-2mm}
}

\foreach \x in {A, ..., Z}{\expandafter\xdef\csname orcid\x\endcsname{\noexpand\href{https://orcid.org/\csname orcidauthor\x\endcsname}
			{\noexpand\orcidicon}}
}
\usepackage{newtxtext,newtxmath}
\usepackage{CJK}
\usepackage{booktabs}
\usepackage[T1]{fontenc}
\usepackage{ae,aecompl}
\usepackage{graphicx}	% Including figure files
\usepackage{amsmath}	% Advanced maths commands
\usepackage{booktabs}
\usepackage{times}
\input{epsf}
\title[WRLO in compact symbiotic systems]{Entering the Wind Roche Lobe Overflow realm in Symbiotic Systems}

\author[R.~F.~Maldonado et al.]{Ra\'{u}l F. Maldonado\thanks{E-mail:\,r.maldonado@irya.unam.mx}$^{1}\orcidA$, Jes\'{u}s~A.~Toal\'{a}$^{1,2}\orcidB$, Emilio~Tejeda$^{3}\orcidC$ and Janis B. Rodr\'{i}guez-Gonz\'{a}lez$^{1}\orcidD$\\
$^{1}$Instituto de Radioastronom\'{i}a y Astrof\'{i}sica, Universidad Nacional Aut\'{o}noma de M\'{e}xico, 58089 Morelia, Michoac\'{a}n, Mexico\\
$^{2}$Facultad de Ciencias de la Tierra y el Espacio, Universidad Aut\'{o}noma de Sinaloa, Josefa Ortiz de Dom\'{i}nguez S/N, Culiac\'{a}n 80040, Sin., Mexico\\
$^{3}$SECIHTI - Instituto de F\'{i}sica y Matem\'{a}ticas, Universidad Michoacana de San Nicol\'{a}s de Hidalgo, Ciudad Universitaria, 58040 Morelia, Mich., Mexico
}

\date{\today}%Accepted XXX. Received YYY; in original form ZZZ}

\pubyear{2025}

\begin{document}

\label{firstpage}
\pagerange{\pageref{firstpage}--\pageref{lastpage}}
\maketitle

\begin{abstract}
We present a suite of dynamical simulations designed to explore the orbital and accretion properties of compact (2--7 AU) symbiotic systems, focusing on wind accretion, drag forces, and tidal interactions. Using three levels of physical complexity, we model systems of accreting white dwarfs (WDs) with masses of 0.7, 1.0, and 1.2 M$_\odot$ orbiting evolving Solar-like stars with 1, 2, and 3 M$_\odot$. We show that systems alternate between standard wind accretion and Wind Roche Lobe Overflow (WRLO) regimes during periods of high mass-loss rate experienced by the donor star (the peak of red giant phase and/or thermal pulses). For some configurations, the standard wind accretion has mass accretion efficiencies similar to those obtained by WRLO regime. Tidal forces play a key role in compact systems, leading to orbital shrinkage and enhanced accretion efficiency. We find that systems with high-mass WDs ($\geq 1$ M$_\odot$) and massive donors (2--3 M$_\odot$) are the only ones to reach the Chandrasekhar limit. Interestingly, the majority of our simulations reach the Roche lobe overflow condition that is not further simulated given the need of more complex hydrodynamical simulations. Our analysis shows that increasing physical realism, by including drag and tides, systematically leads to more compact final orbital configurations. Comparison with compact known symbiotic systems seems to suggest that they are very likely experiencing orbital decay produced by tidal forces. 
\end{abstract}


%%%%%%%%%%%%%%%%%%%%%%%%%%%%%%%%%%%%%%%%%%%%%%%%%%

%%%%%%%%%%%%%%%%% BODY OF PAPER %%%%%%%%%%%%%%%%%%

\section{Introduction}\label{introduction}
\label{sec:intro}

Accretion in binary systems is key to understand a broad variety of astrophysical phenomena and their evolutive paths. In binary systems containing white dwarfs (WDs), accretion is a key physical mechanism to predict possible outcomes as type Ia supernova events, which have cosmological implications \citep{Mukai2017}. Particularly in symbiotic systems, a WD accretes material from a late-type companion that can be a red giant (RG) star or a more evolved star such as an asymptotic giant branch (AGB) star \citep[][]{Mikolajewska2003,Merc2025}.

Mass transfer in binary systems occurs through two primary channels: Roche Lobe Overflow (RLO) and wind accretion \citep{Frank2002}. The former occurs in close systems when the donor star expands beyond its Roche lobe, and its envelope flows through the first Lagrange point into the companion. Studying this process is complex, as the significant mass loss necessitates a self-consistent treatment of stellar evolution, and the transfer itself requires detailed hydrodynamic simulations to accurately capture the binary's response \citep[e.g.][]{Savonije1978,Sawada1992}. In contrast, wind accretion has been successfully modelled using analytic estimates based on the Bondi-Hoyle-Littleton (BHL) framework \citep{Hoyle1939,Bondi1944}. \citet{DavidsonOstriker1973} first applied the BHL model to binary systems where the wind velocity is significantly higher than the orbital velocity. 
Recently, \citet{TejedaToala2025} extended this implementation to handle the more general case of an arbitrary ratio between the stellar wind ($\varv_\mathrm{w}$) and orbital velocity of the system ($\varv_\mathrm{o}$).

The Wind Roche Lobe Overflow (WRLO) scenario was originally introduced to describe accretion in the $\varv_\mathrm{w} \leq \varv_\mathrm{o}$ regime \citep{Mohamed2007}. Here, the wind injection radius of the mass-losing star is comparable to its Roche lobe ($R_\mathrm{Roche}$). Consequently, the wind material is gravitationally bound to the Roche lobe of the mass-losing star and is transferred to the accreting companion through the first Lagrangian point.  Later, simulations adopted the dust condensation radius ($R_\mathrm{cond}$) as a proxy for the wind injection radius. If the $R_\mathrm{cond}\approx R_\mathrm{Roche}$ condition is not met, the classic wind accretion scenario is recovered \citep[see][]{Mohamed2012}.

Based on several numerical simulations presented in \citet{Mohamed2010}, \citet{Abate2013} proposed an empirical formulation to estimate the mass accretion efficiency for the WRLO phase. This resulted in a somewhat complex formulation that depends on the $R_\mathrm{cond}/R_\mathrm{Roche}$ ratio. 
\citet{Abate2013} applied their formulation to study the population of Carbon-enhanced metal-poor stars in the Galactic halo.

The predictions for mass accretion efficiency of the WRLO mechanism established by \citet{Abate2013} have been used to study a variety of exotic phenomena produced in binary systems. For example, the production of blue lurkers and stragglers in star clusters, red giant stars in the thick disc of the Galaxy, the formation of Barium stars, the orbital properties of post-AGB stars \citep[e.g.,][]{Izzard2018,Oomen2018,Sun2024,Krynski2025}, to name a few.

Applications of the WRLO model to symbiotic systems have been presented in \citet{Il2019} and more recently in \citet{Vathachira2025}. The former study investigated the symbiotic system V407 Cyg, a potential Type Ia supernova progenitor due to its high-mass WD, and found that accounting for a WRLO mechanism increases the likelihood of high-mass WDs reaching the Chandrasekhar limit. \citet{Vathachira2025} made an exploration of parameters to identify the conditions under which WRLO should operate as the dominant accretion mechanism in binaries with varying stellar masses and orbital separations during the AGB phase. They suggest that, due to the strong variability associated with thermal pulses in the AGB phase, systems can transition between BHL and WRLO accretion regimes. Both studies conclude that WRLO is highly sensitive to $R_\mathrm{cond}$ and the $R_\mathrm{Roche}$, and that WRLO can significantly enhance mass growth in high-mass WDs, aiding their evolution toward the Chandrasekhar limit.

Here, we try to push forward our understanding of the WRLO phase and its impact on the fate and dynamical evolution of compact symbiotic binary systems in combination with other unexplored physical effects, such is the case of tidal and wind drag forces. The present paper is a continuation of the work presented in \citet{Maldonado2025} (hereinafter Paper I), where stellar evolution models were used in combination with dynamical $N$-body calculations to explore the effects of different wind accretion approaches in the evolution of extended ($a_0 \geq 8$ AU) symbiotic systems in circular orbits. We studied the evolution of the orbital properties of those extended symbiotic systems and the accretion history imprinted in the accretor due to the natural evolution of the stellar wind parameters of the donor star.

This paper is organized as follows. In Section~\ref{sec:methods} we present our methodology, which includes the combination of stellar evolution models with $N$-body simulations. That section describes the physical effects included in the simulations and their setup. Section~\ref{sec:results} presents examples of our results and a discussion is presented in Section~\ref{sec:discussion}. Finally, our summary of conclusions is presented in Section~\ref{sec:summary}.

\section{Methods}
\label{sec:methods}

In this work we use the same stellar evolution models presented in Paper I, which correspond to non-rotating stars with initial masses of 1, 2, and 3~M$_\odot$ computed by the Modules for Experiments in Stellar Astrophysics ({\sc mesa}) code \citep{Paxton2011}. The simulations track the stars from the main sequence phase down to the WD stage, assuming an initial metallicity of $Z = 0.02$ and neglecting magnetic fields. During the red giant branch (RGB) phase, mass loss is implemented using the Reimers   wind efficiency parameter set to 0.5 \citep{Reimers1975}, whilst in the asymptotic giant branch (AGB) phase we adopt the mass loss efficiency parameter of 0.1  \citep{Bloecker1995}. Following Paper I, the mass-loss rate evolution is directly taken from the stellar evolution models, whilst the terminal velocity of the stellar wind $\varv_\mathrm{w}$ is estimated using the empirical scaling proposed by \citet{Verbena2011}:
\begin{equation}
\varv_\mathrm{w} = 0.05 \left( \frac{L_1}{\mathrm{L}_\odot} \cdot \frac{\mathrm{M}_\odot}{m_1}\right)^{0.57}  {\mathrm{km}~\mathrm{s}^{-1}},
\label{eq:verbena}
\end{equation}
where $L_1$ is the luminosity of the mass-losing star of mass $m_1$. 
The calculated wind velocity $\varv_\mathrm{w}$, as well as the mass-loss rate $\dot{M}_\mathrm{w}$ of our {\sc mesa} models are shown in Appendix \ref{app:vwind_mdot}.

To carry out the dynamical evolution of our evolving symbiotic systems we use the $N$-body code {\sc rebound} \citep{Rein2011,Rein2012} together with the {\sc reboundx} extension package \citep{Tamayo2019}. 
The simulations are resolved using the IAS15 integrator \citep{Rein2015}, a 15th-order Gauss–Radau scheme with adaptive time-stepping, optimized for high-precision integrations.  
We model symbiotic binary systems in circular orbits consisting of evolving donor stars with initial masses of $m_{1,0} = 1$, 2, and 3~M$_\odot$, and accreting WD companions with masses of $m_{2,0} = 0.7$, 1.0, and 1.2 M$_\odot$. In this paper we explore simulations adopting initial orbital separations of $a_0 = 2$, 3, 4, 5, 6, and 7 AU. These system combinations result in initial orbital periods with values between 2 and 15 yr.

\subsection{Mass accretion channels}

All models start in a wind accretion regime that is initially modelled following the modified BHL accretion scheme proposed by \citet{TejedaToala2025}. The wind accretion efficiency, defined as the ratio of the mass accretion rate over the mass loss rate from the donor star,  $\eta=\dot{M}_\mathrm{acc}/\dot{M}_\mathrm{w}$, can be expressed for the circular case as:
\begin{equation}
\eta = \left(\frac{q}{1+\varw^2}\right)^2,
\label{eq:TejedaToala}
\end{equation}    
\noindent where 
\begin{equation}
    q = \frac{m_2}{m_1 + m_2}
\end{equation}
\noindent is the dimensionless mass ratio and 
\begin{equation}
    \varw = \frac{\varv_\mathrm{w}}{\varv_\mathrm{o}}    
    \label{eq:w}
\end{equation}
\noindent is the dimensionless velocity ratio. The orbital velocity is given by 
\begin{equation}
    \varv_\mathrm{o} = \sqrt{\frac{G (m_1 + m_2)}{a}},
    \label{eq:orb_vel}
\end{equation} with $a$ being the instantaneous orbital separation and $G$ the gravitational constant.

In addition to the standard wind accretion model, we allow the simulated systems to experience the WRLO accretion mechanism. The evolution of the mass losing star $m_1$ during the RGB and AGB phases naturally increases its radius $R_1$ and so does the dust condensation radius $R_\mathrm{cond}$. In this work we approximate the condensation radius following \citet{Hofner2007}:
\begin{equation}
    R_\mathrm{cond} = \frac{R_1}{2} \left(\frac{T_1}{T_\mathrm{cond}} \right)^{2},
\label{eq:rcond}    
\end{equation}
\noindent where $T_1$ is the effective temperature of the donor star and $T_\mathrm{cond} = 1500$ K is the adopted dust condensation temperature, appropriate for amorphous carbon dust \citep{Whittet2022}.

The Roche lobe radius $R_\mathrm{Roche}$ is computed using \citet[][]{Eggleton1983}'s approximation, where
\begin{equation}
    \frac{R_\mathrm{Roche}}{a} = \frac{0.49~\delta^{2/3} }{0.6~\delta^{2/3} + \ln(1 + \delta^{1/3})},
\label{eq:RL}    
\end{equation}
\noindent with $\delta$ as the donor versus accretor mass ratio, that is, $\delta = m_1/m_2$.

If the condition $R_\mathrm{cond}  \geq R_\mathrm{Roche}$ is met during the evolution of our systems, the simulation transitions to the WRLO accretion regime. In this case, the accretion efficiency is updated to $\eta_\mathrm{WRLO}$, following the proposed formulation of \citet{Abate2013}, where
\begin{equation}
\begin{split}
\eta_{\mathrm{WRLO}} = \min \bigg\{ \frac{25}{9} q_\mathrm{m}^2 \bigg[ 
    -0.284\left(\frac{R_\mathrm{cond}}{R_\mathrm{Roche}}\right)^2 
    + 0.918\frac{R_\mathrm{cond}}{R_\mathrm{Roche}} \\
    - 0.234 \bigg],\ 0.5 \bigg\}.
\end{split}
\label{eq:abate}
\end{equation}
\noindent Here, $q_\mathrm{m}= \delta^{-1} = m_2/m_1$ and the upper limit of 0.5 reflects the maximum accretion efficiency reported by numerical simulations presented in \citet{Mohamed2010}.

\subsection{Extra physical mechanisms}

\subsubsection{Wind drag force}

We define the wind drag force following the formalism of \citet[][]{Hoyle1939} as
\begin{equation}
\vec{F}_\mathrm{drag}=\dot{M}_\mathrm{acc}~\vec{\varv}_\mathrm{rel},
\end{equation}
\noindent where $\vec{\varv}_\mathrm{rel}$ denotes the relative velocity between the accreting body and the stellar wind, and $\dot{M}_\mathrm{acc}$ is the mass accretion rate given by $\dot{M}_\mathrm{acc} = \eta~\dot{M}_\mathrm{w}$, where $\eta$ is the accretion efficiency, determined either by Eq.~(\ref{eq:TejedaToala}) or (\ref{eq:abate}), depending on the experienced wind accretion regime. This drag force is implemented as an external force within the {\sc reboundx} framework.

\subsubsection{Tidal interactions}

To model the tidal interactions, we implement the weak friction model of \citet{Hut1981} using the \texttt{tides$\_$constant$\_$time$\_$lag} module in {\sc reboundx} \citep[see][]{Baronett2022}. In this framework, the gravitational pull of the secondary generates a tidal bulge on the primary star. Because the primary’s response is delayed by a constant time lag, the bulge does not align perfectly with the axis connecting the centers of mass, leading to a torque. This torque dissipates orbital energy and gradually modifies both the orbital configuration and the rotational state of the system.

The constant time lag tidal model is governed by two principal parameters. The first is the time lag itself, a non-zero constant defined by
\begin{equation} 
\tau(t) = \frac{2R_1^3}{G m_1 t_\mathrm{f}}. 
\label{eqtau} 
\end{equation}
\noindent The quantity $t_\mathrm{f}$, known as the convective friction time, represents the timescale over which the star's convective envelope responds to tidal forcing, effectively controlling the phase lag of the tidal bulge.

The convective friction time evolves with the stellar structure and is expressed as \citep[see][]{Zahn1989,Schroder2008}
\begin{equation} 
t_\mathrm{f}(t) = \left(\frac{m_1 R_1^2}{L_1}\right)^{1/3}, 
\label{eqtf} \end{equation}
\noindent where $L_1$ is the luminosity of the mass losing star. This formulation captures the dependence of tidal dissipation efficiency on the internal properties of the star.

The second fundamental parameter in the tidal model is the second-order tidal Love number, $k_2$, which quantifies the primary star’s deformability in response to the tidal potential. This parameter is highly sensitive to the star’s internal density profile and thus evolves with its structure.

\begin{figure}
\begin{center}
\includegraphics[width=\linewidth]{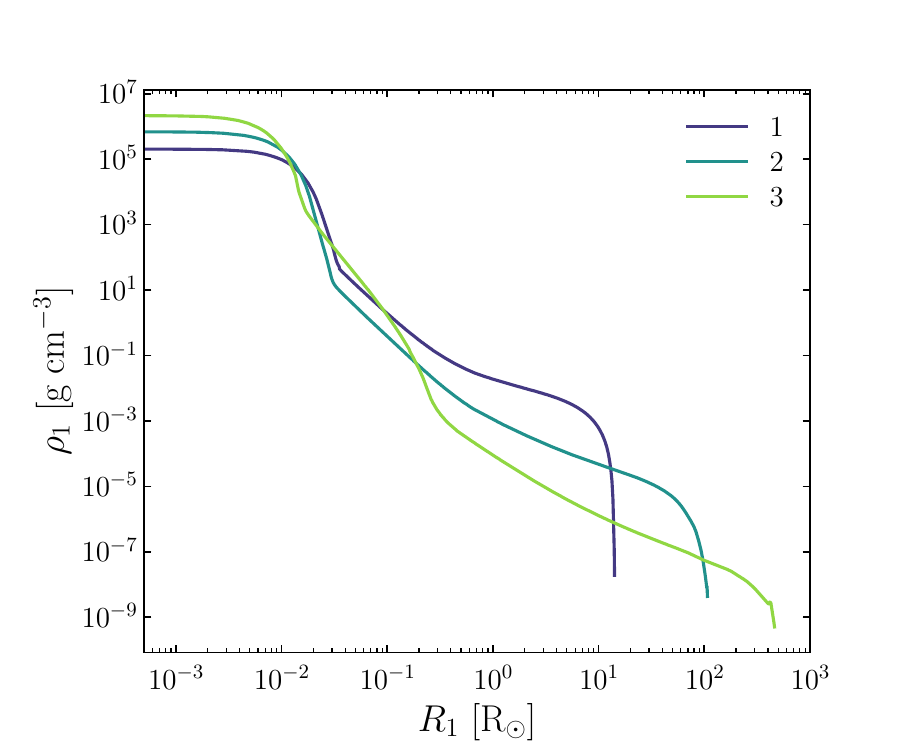}
\end{center}
\caption{Density profiles of the donor star with $m_{1,0}=2$ M$_\odot$ as a function of stellar radius at selected evolutionary stages. The profiles correspond to (1) the beginning of the RG phase, (2) the maximum expansion at the RGB phase, and (3) the largest expansion achieved during the TPAGB phase.} 
\label{fig:density}
\end{figure}

The Love number $k_2$ can be determined using the method described by \citet{Batygin2009} and \citet{Becker2013}, where
\begin{equation} 
k_2 = \frac{3 - \zeta(R_1)}{2 + \zeta(R_1)}. 
\label{eq:loven} 
\end{equation}
\noindent Here $\zeta(r)$ is a dimensionless function derived by solving the differential equation
\begin{equation} 
r \frac{d\zeta(r)}{dr} + \zeta(r)^2 - 6 + 6 \frac{\rho_1(r)}{\rho_{\mathrm{1,avg}}(r)} [\zeta(r) + 1] = 0, 
\label{eq:loven2} 
\end{equation}
\noindent integrated from the centre of the star ($r = 0$) to the surface ($r = R_1$). Here, $\rho_1(r)$ is the local density at radius $r$, and $\rho_{\mathrm{1,avg}}(r)$ is the mean density enclosed within that radius. The density profile of the star at each time is obtained from the {\sc mesa} models. To illustrate the changes in density structure experienced by an evolving star, we show in Fig.~\ref{fig:density} examples of the density profile of the $m_{1,0}= 2$ M$_\odot$ model used in this work.

It is worth noticing that we do not evolve the spin of either star or include rotational feedback. This simplification isolates the role of tides on orbital decay, without the added complexity of spin–orbit coupling or changes in moment of inertia. In practice, this assumes tides always act at maximum efficiency, as if the donor rotates more slowly than the orbit \citep{Zahn1977,Zahn1989,Hut1981}. For convective-envelope giants, synchronization is generally achieved on short timescales \citep{Verbunt1995,Nordhaus2013}. However, subsequent stellar expansion, for example during the thermal pulses, can reintroduce asynchronicity allowing tides to act again. Including such effects will be an important avenue for future studies in evolving symbiotic systems.

\begin{figure*}
\begin{center}
\includegraphics[width=0.95\linewidth]{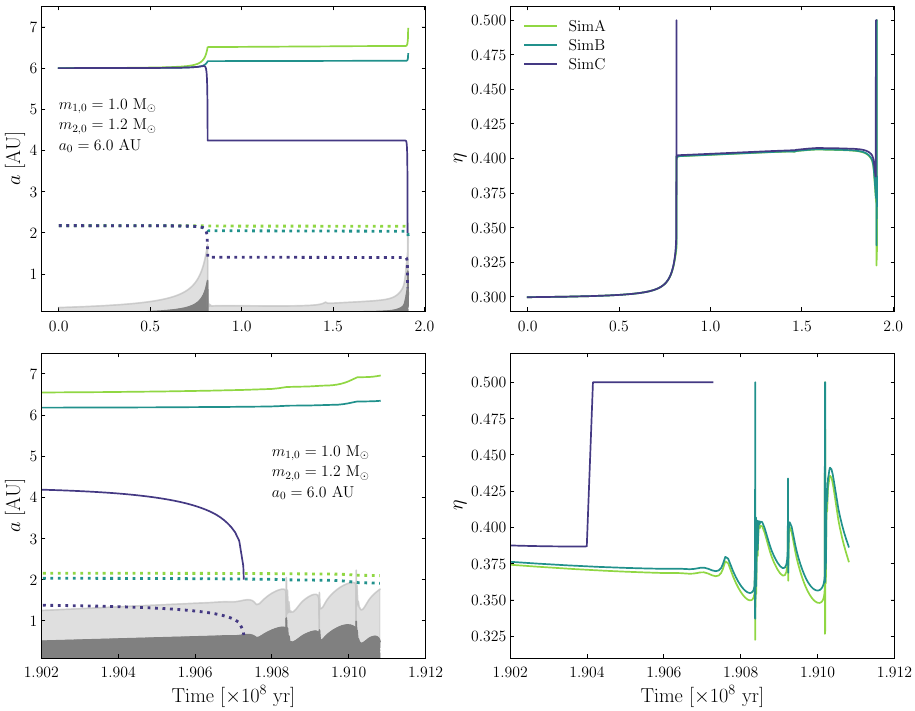}
\end{center}
\caption{Semimajor axis evolution (left panels) and accretion efficiency over time (right panels) for a symbiotic binary system with initial masses $m_{1,0}=1$ $M_\odot$, $m_{2,0}=1.2$ M$_\odot$ and an initial separation $a_0=6$ AU. The top panels show the entire duration of the integration, while the bottom panels focus on the TPAGB phase. Different colours correspond to simulations with varying levels of physical complexity, as described in Section~\ref{sec:2.3}. In the left panels, the dotted lines trace the evolution of the donor star's Roche lobe radius. The light gray shaded region denotes the dust condensation radius ($R_\mathrm{cond}$), and the darker gray area marks the stellar surface radius ($R_1$). } 
\label{fig:a_eta}
\end{figure*}

Because key stellar properties, including $m_1$, $m_2$, $R_1$, $L_1$, $\varv_{\mathrm{w}}$, $\varv_{\mathrm{o}}$, $\dot{M}_{\mathrm{w}}$, $\tau$, and $k_2$, evolve throughout the star’s lifetime, these quantities are continuously updated during the simulation. At each time step, values are interpolated from the stellar evolution model to reflect the current state of the star.

\subsection{The setup}
\label{sec:2.3}

We carry out three different sets of simulations with increasing levels of physical complexity:

\noindent \underline{SimA:} Wind accretion only. In these simulations the wind accretion scheme of \citet{TejedaToala2025} is alternated with the WRLO scheme from \citet{Abate2013}. The latter is activated when the condition $R_\mathrm{cond} \geq R_\mathrm{Roche}$ is met.

\noindent \underline{SimB:} Wind accretion + wind drag force. Same as SimA simulations but the effects of the wind drag force are also considered.

\noindent \underline{SimC:} Wind accretion + wind drag force + tidal interactions. All physical mechanisms are included in the calculations. In these simulations, both the donor star and the accretor are assumed to be non-rotating.

As in Paper I, the simulations of the evolving symbiotic systems are tracked from the onset of the RGB phase through to the end of the thermally pulsing AGB (TPAGB) phase, defined as the point when the star's surface temperature increases to $\log_{10}(T_\mathrm{eff}/\mathrm{K}) = 3.6$. 
However, if the accreting companion reaches the Chandrasekhar mass limit of 1.4~M$_\odot$, the threshold for triggering a Type Ia supernova, the simulation is terminated. Furthermore, during the RGB or TPAGB phase, the donor star can expand significantly potentially filling its Roche lobe, that is, $R_1 = R_\mathrm{Roche}$. Since the onset of Roche lobe overflow (RLO) is not modelled in this work, the simulation is stopped if this condition is met. Additionally, following the prescription of \citet{Abate2013}, the WRLO accretion efficiency  $\eta_\mathrm{WRLO}$ approaches zero as the ratio $\frac{R_\mathrm{cond}}{R_\mathrm{Roche}}$ tends to 2.951. At such small Roche lobe radii, a significant fraction of the wind is expected to escape through the $L_2$ and $L_3$ Lagrangian points. This situation arises particularly in simulations where tidal interactions cause orbital shrinking and reduce $R_\mathrm{Roche}$. This situation represents another stopping condition in our simulated systems.   

We conducted a total of 162 {\it N}-body simulations of compact symbiotic binary systems in circular orbits consisting in a combination of a donor star with initial masses of $m_{1,0} = 1$, 2, and 3 M$_\odot$, and a WD accretor with masses of $m_{2,0} = 0.7$, 1.0, and 1.2 M$_\odot$ with initial semimajor axis between $a_0=2$ and $a_0=7$ AU. For simplicity, examples of the results will be described only using simulations with $a_0 = 6$ AU, but further discussion includes the properties of all numerical simulations as a sample.

\section{Results}
\label{sec:results}

\begin{figure*}
\begin{center}
\includegraphics[width=0.95\linewidth]{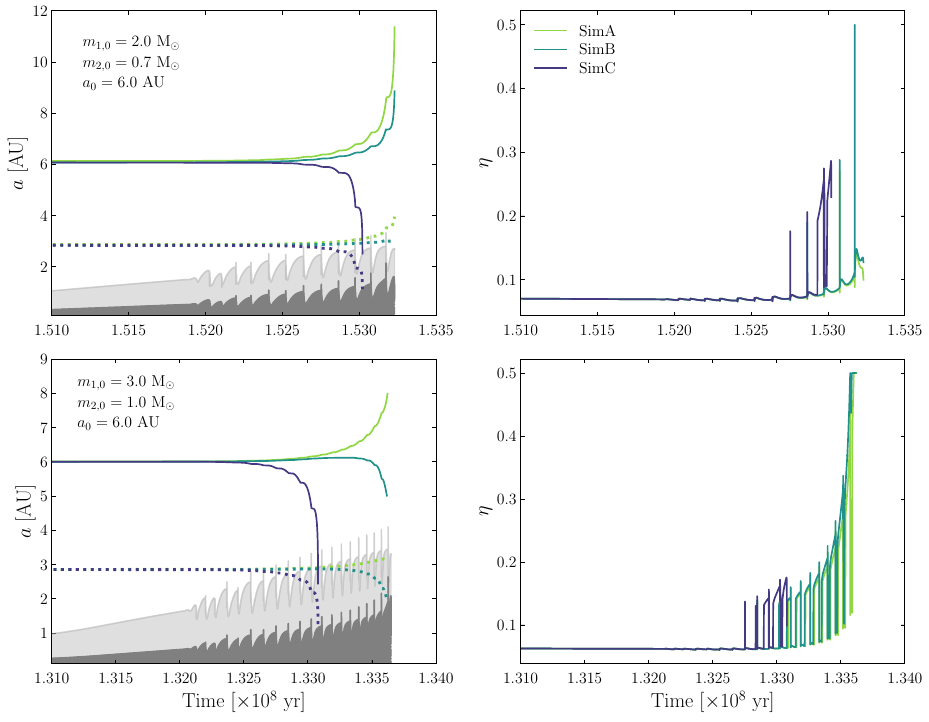}
\end{center}
\caption{Similar to Figure \ref{fig:a_eta} but illustrating the evolution of a symbiotic system during the TPAGB phase with $m_{1,0}=2$ M$_\odot$ (top panels) and $m_{3,0}=3$ M$_\odot$ (bottom panels). } 
\label{fig:a_eta2}
\end{figure*}

\begin{table}
\begin{center}
\caption{Final masses at the end of each set of simulations. The subindexes, A, B, and C refer to the three sets of simulations performed in this study, respectively. The three final columns describes the final destination or the stopping condition of the simulation: $E$ - the simulation ends without interruption after the TPAGB phase, $C$ - the accretor reaches the Chandrasekhar limit, and $R$ - the donor star fills its Roche lobe radius ($R_1=R_\mathrm{roche}$).}
%\footnotesize
\setlength{\tabcolsep}{0.75\tabcolsep}  
%\scriptsize
\begin{tabular}{lcccccccc}
\hline
$m_{1,0}$    & $m_{2,0}$   &  $a_0$   & $m_\mathrm{2F,A}$ & $m_\mathrm{2F,B}$ & $m_\mathrm{2F,B}$ & $D_\mathrm{A}$ & $D_\mathrm{B}$ & $D_\mathrm{C}$ \\
(M$_\odot$)  & (M$_\odot$) &  (AU)          &  (M$_\odot$) &  (M$_\odot$)&  (M$_\odot$) & & & \\
(1) & (2) & (3) & (4) & (5) & (6) & (7) & (8) & (9) \\
\hline 
1.0 &	0.7 &    2.0   & 0.803 & 0.783 & 0.709 & $R$ & $R$ & $R$\\
    &	    &    3.0   & 0.887 & 0.829 & 0.720 & $E$ & $R$ & $R$\\
    &       &    4.0   & 0.826 & 0.849 & 0.733 & $E$ & $E$ & $R$\\
    &       &    5.0   & 0.795 & 0.797 & 0.747 & $E$ & $E$ & $R$\\
    &       &    6.0   & 0.791 & 0.793 & 0.794 & $E$ & $E$ & $E$\\
    &       &    7.0   & 0.788 & 0.790 & 0.790 & $E$ & $E$ & $E$\\

 &	 1.0   &   2.0   & 1.093 & 1.083 & 1.011 & $R$ & $R$ & $R$ \\
 &	       &   3.0   & 1.204 & 1.143 & 1.024 & $E$ & $R$ & $R$ \\
 &          &   4.0   & 1.177 & 1.185 & 1.039 & $E$ & $E$ & $R$ \\ 
 &          &   5.0   & 1.145 & 1.154 & 1.058 & $E$ & $E$ & $R$ \\
 &          &   6.0   & 1.132 & 1.134 & 1.153 & $E$ & $E$ & $E$ \\
 &          &   7.0   & 1.128 & 1.130 & 1.131 & $E$ & $E$ & $E$ \\

 &   1.2   & 2.0  & 1.289  & 1.282  & 1.212  & $R$  & $R$ & $R$ \\
 &         & 3.0  & 1.349  & 1.349  & 1.225  & $R$  & $R$ & $R$ \\
 &         & 4.0  & 1.395  & 1.399  & 1.243  & $E$  & $E$ & $R$ \\
 &         & 5.0  & 1.370  & 1.376  & 1.262  & $E$  & $E$ & $R$ \\
 &         & 6.0  & 1.356  & 1.357  & 1.305  & $E$  & $E$ & $R$ \\
 &         & 7.0  & 1.351  & 1.353  & 1.354  & $E$  & $E$ & $E$ \\
\hline
2.0 &	0.7  &   2.0   & 0.712 & 0.712 & 0.702 & $R$ & $R$ & $R$ \\
    &	     &   3.0   & 0.767 & 0.741 & 0.704 & $R$ & $R$ & $R$ \\
    &        &   4.0   & 0.990 & 0.819 & 0.707 & $R$ & $R$ & $R$ \\
    &	     &   5.0   & 1.175 & 0.967 & 0.714 & $E$ & $R$ & $R$ \\ 
    &        &   6.0   & 0.836 & 0.854 & 0.740 & $E$ & $E$ & $R$ \\
    &        &   7.0   & 0.827 & 0.835 & 0.926 & $E$ & $E$ & $E$ \\

 &	   1.0  &    2.0   & 1.023 & 1.020 & 1.003 & $R$ & $R$ & $R$\\
 &	        &    3.0   & 1.086 & 1.055 & 1.007 & $R$ & $R$ & $R$\\
 &           &    4.0   & 1.388 & 1.128 & 1.010 & $R$ & $R$ & $R$\\
 &	        &    5.0   & 1.400 & 1.353 & 1.020 & $C$ & $R$ & $R$\\
 &           &    6.0   & 1.234 & 1.277 & 1.046 & $E$ & $R$ & $R$\\
 &           &    7.0   & 1.211 & 1.231 & 1.119 & $E$ & $E$ & $R$\\

 &   1.2   & 2.0  & 1.227  & 1.227  & 1.204  & $R$  & $R$ & $R$ \\
 &         & 3.0  & 1.269  & 1.249  & 1.208  & $R$  & $R$ & $R$ \\
 &         & 4.0  & 1.400  & 1.342  & 1.212  & $C$  & $R$ & $R$ \\
 &         & 5.0  & 1.400  & 1.400  & 1.223  & $C$  & $C$ & $R$ \\
 &         & 6.0  & 1.400  & 1.400  & 1.250  & $C$  & $C$ & $R$ \\
 &         & 7.0  & 1.400  & 1.400  & 1.304  & $C$  & $C$ & $R$ \\

\hline
3.0 &	0.7 &    2.0   & 0.701 & 0.701 & 0.700 & $R$ & $R$ & $R$ \\
    &	    &    3.0   & 0.720 & 0.715 & 0.700 & $R$ & $R$ & $R$\\
    &       &    4.0   & 1.039 & 0.776 & 0.701 & $R$ & $R$ & $R$\\
    &	    &    5.0   & 1.362 & 0.982 & 0.703 & $E$ & $R$ & $R$\\
    &       &    6.0   & 0.842 & 0.946 & 0.710 & $E$ & $R$ & $R$\\
    &       &    7.0   & 0.829 & 0.842 & 0.737 & $E$ & $E$ & $R$\\

 &	  1.0   &    2.0   & 1.002 & 1.002 & 1.001 & $R$ & $R$ & $R$\\
 &	        &    3.0   & 1.023 & 1.017 & 1.001 & $R$ & $R$ & $R$\\
 &           &    4.0   & 1.276 & 1.109 & 1.001 & $R$ & $R$ & $R$\\
 &	        &    5.0   & 1.400 & 1.265 & 1.004 & $C$ & $R$ & $R$\\
 &           &    6.0   & 1.400 & 1.400 & 1.013 & $C$ & $C$ & $R$\\
 &           &    7.0   & 1.234 & 1.238 & 1.038 & $E$ & $E$ & $R$\\

 &   1.2   & 2.0  & 1.202  & 1.202  & 1.201 & $R$ & $R$  & $R$ \\
 &         & 3.0  & 1.224  & 1.224  & 1.201 & $R$ & $R$  & $R$ \\
 &         & 4.0  & 1.400  & 1.313  & 1.202 & $C$ & $R$  & $R$ \\
 &         & 5.0  & 1.400  & 1.400  & 1.205 & $C$ & $C$  & $R$ \\
 &         & 6.0  & 1.400  & 1.400  & 1.215 & $C$ & $C$  & $R$ \\
 &         & 7.0  & 1.400  & 1.400  & 1.241 & $C$ & $C$  & $R$ \\

\hline
\end{tabular}
\label{tab:times}
\end{center}
\end{table}

Fig.~\ref{fig:a_eta} presents the evolution of the semimajor axis $a$ and mass accretion efficiency $\eta$ for a symbiotic binary systems with $m_{1,0}=1$ M$_\odot$ and $m_{2,0}=1.2$ M$_\odot$ at an initial orbital separation of $a_0=6$ AU. Results from SimA, SimB, and SimC are presented in solid lines. 

The evolution of the binary separation $a$ in SimA and SimB is very similar, leading to expanded orbits at the end of their evolution. The expansion is driven by the substantial mass lost from the system, which leads to a natural reduction in angular momentum (see Fig.\ref{fig:a_eta}, top-left). This effect is less pronounced in SimB, but it seems that the inclusion of wind drag force does not impact dramatically the evolution of the symbiotic system with such initial configuration compared to SimA. These two cases, reach the end of the simulation (end of the TBPAG phase).

The most distinct case for this selected configuration is SimC, where tidal interactions cause the orbit to shrink during the initial significant expansion of the mass-losing star’s radius at the end of the RGB phase (see Fig.~\ref{fig:a_eta}, top left). As a result, the orbit contracts to $a = 4.1$ AU and remains at this configuration until the star enters the TPAGB phase, during which its radius increases once again and the role of tides is regained. SimC stops because the radius of the mass losing star reaches the Roche lobe radius ($R_1 = R_\mathrm{Roche}$; Fig.~\ref{fig:a_eta}, bottom left), a condition that should give rise to a RLO phase, not modelled in the present work. SimC stops about $3.6\times10^{5}$ yr before the end of the TPAGB phase.

Fig.~\ref{fig:a_eta} also shows the evolution of the Roche lobe radius ($R_\mathrm{Roche}$, dotted lines), the donor star's radius ($R_1$, dark grey shaded region), and the dust condensation radius ($R_\mathrm{cond}$, light grey shaded region) which can be used to assess the times in which the simulations entered the WRLO or the moment they stop given the RLO condition. For example, SimC experiences the WRLO during two times, one at the RGB phase and another just before the onset of the TPAGB phase (Fig.~\ref{fig:a_eta}, left panels). SimA and SimB experience the WRLO phase in very short periods during the evolution of the TPAGB phase, specifically, at extreme mass-lost rates periods produced by thermal pulses of the donor star. Fig.~\ref{fig:eta_wrlo} in Appendix~\ref{app:WRLO} illustrates further the times where the WRLO phase in these simulations is active.

Other examples are presented in Fig.~\ref{fig:a_eta2}. The top panels show simulations corresponding to evolving symbiotic system with $m_{1,0} = 2.0$ M$_\odot$ and $m_{2,0} = 0.7$ M$_\odot$ while the bottom panels correspond to calculations with $m_{1,0} = 3.0$ M$_\odot$ and $m_{2,0} = 1.0$ M$_\odot$. In both cases we only show the evolution during the TPAGB phase given that these simulations are not dramatically affected by any physical mechanism in the previous evolutive phases\footnote{We note that the stellar evolution models of the 2 and 3 M$_\odot$ mass stars used here, originally presented in Paper I, do not produce a considerable increase of the donor star's radius during the earlier RGB phase.}.

The models presented in the top panels of Fig.~\ref{fig:a_eta2} behave very similar to those illustrated in Fig.~\ref{fig:a_eta}, with calculations of SimA and SimB resulting in expanded orbits, also reaching the end of the TPAGB phase. Under this configuration, SimC is stopped after $R_1=R_\mathrm{Roche}$, about $2.1\times10^{5}$ yr before the end of the TPAGB phase and no further evolution is followed. We highlight the WRLO phase of these models in the left panels of Fig.~\ref{fig:eta_wrlo2}.

Finally, the bottom panels of Fig.~\ref{fig:a_eta2} show that in the case of the $m_{1,0} = 3.0$ M$_\odot$ and $m_{2,0}=1.0$ M$_\odot$ simulations, SimA and SimC behave similarly to the previous discussed cases. SimA expands its orbit, while SimC is terminated at the RLO condition $5.7\times10^{5}$ yr before the end of the TPAGB phase. SimB does not end up in an expanded orbit as in the previous cases. The combination of the drag force with the rapid expansion of the radius of the mass losing star cause the shrinkage of the orbit. Here, SimA and SimB are terminated $2.4\times 10^4$ yr and $3.3\times10^4$ yr before the end of the TPAGB phase because their accretors reach the Chandrasekhar limit, respectively. The WRLO phase in these simulations is emphasized in the right panels of Fig. \ref{fig:eta_wrlo2}.

The examples shown in Fig.~\ref{fig:a_eta} and \ref{fig:a_eta2} broadly represent most of the outcomes obtained from all of our 162 simulations. They confirm previous suggestions that symbiotic systems can experience different wind accretion regimes \citep[e.g.,][]{Vathachira2025}. In some cases going from standard wind accretion to the WRLO regime and back. In contrast, we note that the systems modelled in Paper I, which were initialized with $a_0 > 8$ AU evolved only through the wind accretion regime expanding their orbits.

As a summary, Table~\ref{tab:times} presents the final destination of all of the 162 simulations. The first column lists the initial mass of the donor star ($m_{1,0}$), the second column shows the initial mass of the accreting WD ($m_{2,0}$), and the third column provides the initial semimajor axis ($a_0$). Columns 4--6 present the final mass of the accretor resulted from simulations incorporating increasing levels of physical complexity, labelled with subscripts A, B, and C ($m_\mathrm{2F,A}$, $m_\mathrm{2F,B}$, and $m_\mathrm{2F,C}$). The final three columns (columns 7--9) indicate the simulation destiny (e.g., stopping condition): ($E$) denotes that the calculation terminated at the end of the TPAGB phase, ($C$) is used to denote cases in which the accretor reached the Chandrasekhar limit, and ($R$) is for those cases in that reached the RLO condition.

It is evident that additional physical processes play a key role in determining system outcomes. For example, SimB cases, which include wind drag forces, consistently show slightly higher accreted mass than those with accretion alone (SimA), best seen in cases of simulations that completed their evolution up to the TPAGB phase. The explanation is that drag forces promotes orbital decay, increasing the orbital velocity (see Eq.~\ref{eq:orb_vel}) and consequently increasing the mass accretion efficiency (see Eq.~\ref{eq:TejedaToala} and \ref{eq:w}). However, this same orbital decay also raises the probability of RLO, leading to premature termination of the simulation. Approximately  56 per cent of SimB simulations end in RLO compared to only 37 per cent in the SimA set.

Simulations including tidal forces (SimC) are impacted by strong inward orbital migration, shrinking the donor’s Roche lobe and increasing the likelihood of experiencing the RLO. This is especially effective during the TPAGB phase, when the donor’s radius expands rapidly. About 90 per cent of SimC simulations are interrupted when reaching the RLO condition during the TPAGB (see column 9 of Table~\ref{tab:times}), before the accretor can reach the Chandrasekhar limit. Interestingly, we note that \citet{Munari2025} have recently suggested that the RG star in the symbiotic recurrent nova T CrB is filling its Roche lobe. This symbiotic system has an orbital period of 227~d, which agrees with our predictions that compact systems should be accreting through RLO mechanism. 
However, in evolving mass losing stars where the radius growth during the RGB phase is significative, such is the case of our $m_{1,0}=1.0$ M$_\odot$ model, the WRLO is also possible.

\begin{figure}
\begin{center}
\includegraphics[width=\linewidth]{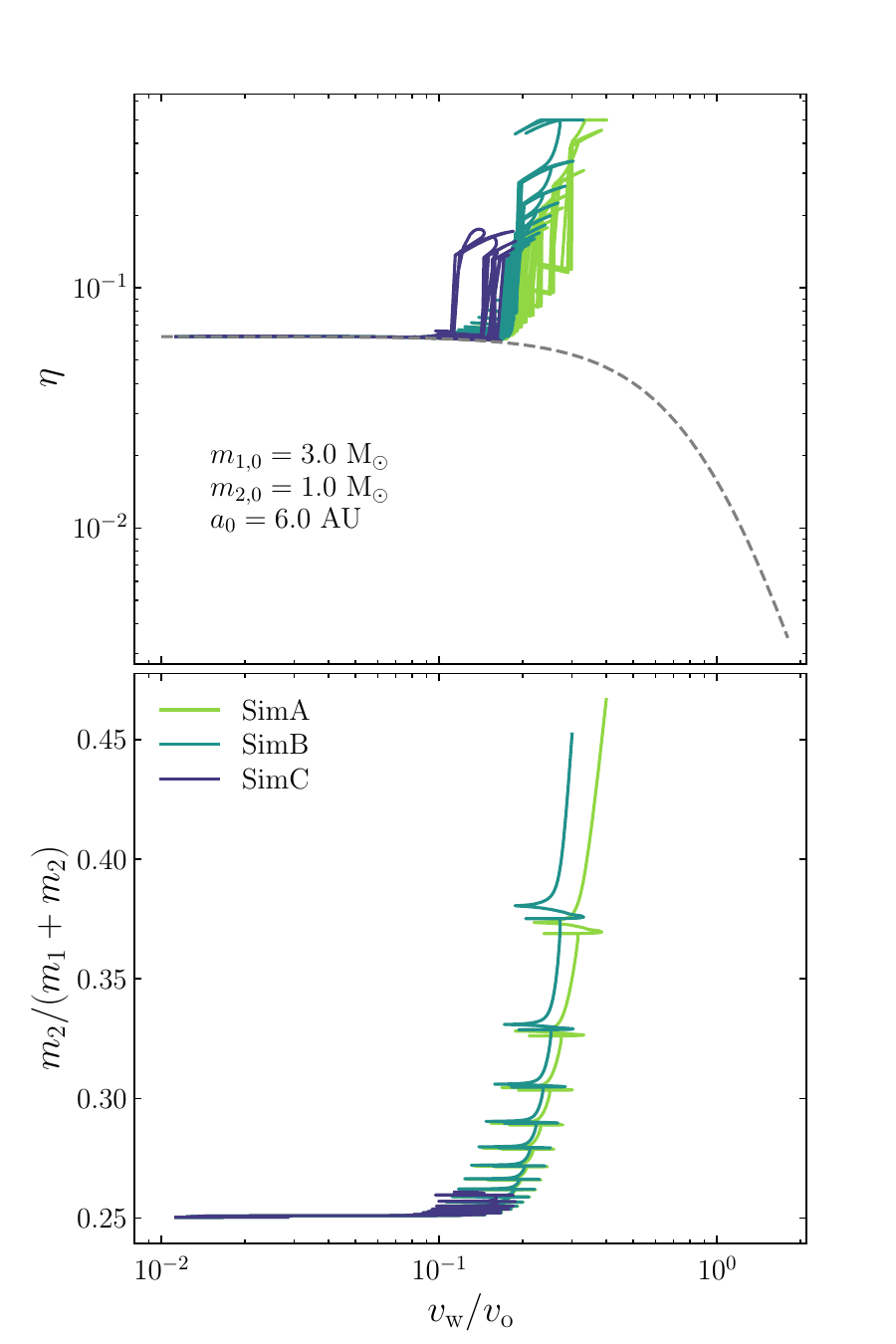}
\end{center}
\caption{Evolution of the accretion efficiency ($\eta$ - top panel) and dimensionless mass ratio ($q=m_2/(m_1+m_2)$ - bottom panel) as a function of the dimensionless velocity parameter  ($\varw=\varv_\mathrm{w}/\varv_\mathrm{o}$) for models with $m_{1,0}= 3.0$ M$_\odot$, $m_{2,0} = 1.0$ M$_\odot$, and $a_0=6$ AU. The dashed line represents the accretion efficiency predicted by \citet{TejedaToala2025} for the adopted initial configuration of the systems.}
\label{fig:eta_w}
\end{figure}

For comparison with previous predictions of extended wind accreting symbiotic systems of Paper I, in Fig.~\ref{fig:eta_w} we present the evolution of the wind accretion efficiency $\eta$ (top) and the dimensionless mass ratio $q = m_2 / (m_1 + m_2)$ (bottom) as functions of the wind velocity ratio $\varw = \varv_\mathrm{w} / \varv_\mathrm{o}$, for models with $m_{1,0} = 3.0$ M$_\odot$ and $m_{2,0}=1.0$ M$_\odot$. We find that in the three cases (SimA, SimB, and SimC) the mass accretion efficiency increases considerably, while in the extended symbiotic systems modelled in Paper I they all evolve into the lower mass accretion range defined by $\varw > 1$ and illustrated by the dashed line in Fig.~\ref{fig:eta_w} (see for comparison figure 9 of Paper I). In general, all the simulations studied in the present work evolve in the $\varw < 1$ range, placing the systems firmly outside the regime where the classical BHL accretion model is valid, reinforcing the need of applying the modified wind accretion scheme of \citet[][]{TejedaToala2025}.

\begin{figure}
\includegraphics[width=\linewidth]{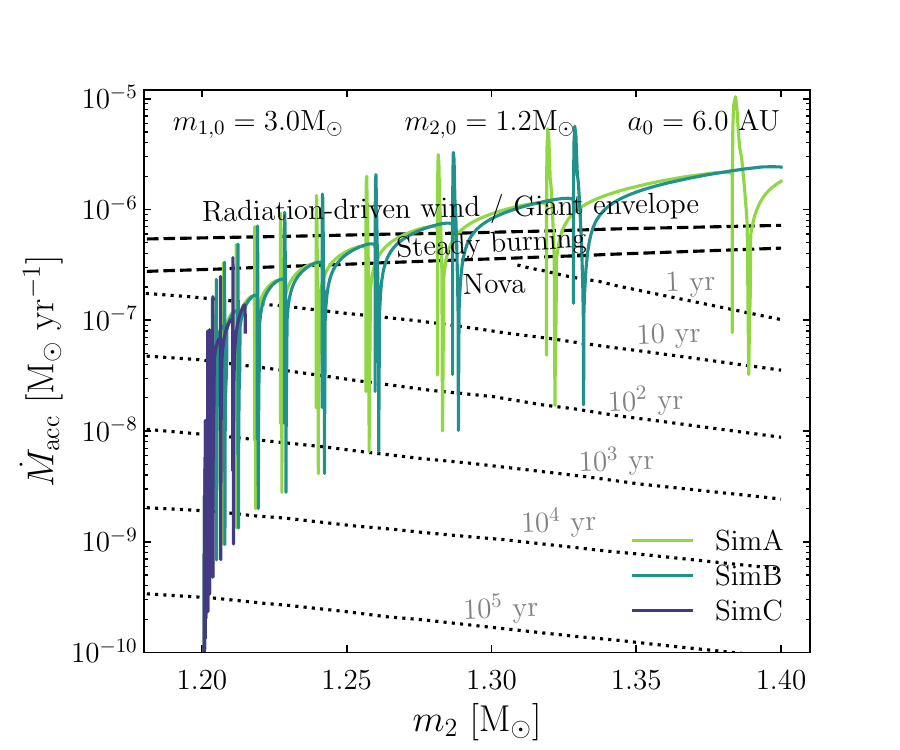}
\caption{Mass accretion rate $\dot{M}_\mathrm{acc}$ as a function of the accretor's mass for simulations with $m_{1,0}= 3.0$ M$_\odot$, $m_{2,0} = 1.2$ M$_\odot$, and $a=6$ AU. Different colours refer to levels of physical complexity in the simulations. This figure was adapted from \citet{Chomiuk2021} which adapted it from results presented in \citet{Wolf2013}. Dotted lines show constant nova recurrence times. The dashed lines show the limits between the three
accretion regimes (nova recurrence, steady burning, and radiation-driven wind). }
\label{fig:mac_mf}
\end{figure}

Both $\eta$ and $q$ are observed to increase significantly, closely tracking the pulsation behaviour of the donor star during the TPAGB phase. 
A particularly striking behaviour is seen in the evolution of $\eta$, which does not increase smoothly but instead shows sharp jumps and drops, associated with the system entering and exiting the WRLO regime (see Fig.~\ref{fig:eta_wrlo2}). These fluctuations are driven by the thermal pulses during the TPAGB phase, which modulate the efficiency of mass transfer.

Finally, in Fig.~\ref{fig:mac_mf} we show examples of the evolution of our simulations in the $\dot{M}_\mathrm{acc}$ versus $m_{2}$ diagram. The simulations presented here have a very similar behaviour than their extended orbit counterparts presented in Paper I. A single evolving system transitions between the different accretion regimes experienced by accreting WDs: nova recurrence, steady burning, and the radiation-driven wind \citep[see][]{Wolf2013,Chomiuk2021}.

\section{Discussion}
\label{sec:discussion}

In this study, we simulate compact symbiotic binaries under various physical configurations, incorporating wind accretion, wind drag, and tidal interactions. Our results support previous suggestions that a single evolving symbiotic system can transition between the standard wind accretion and WRLO regimes, particularly during periods of high mass-loss rate experienced by the donor star (the peak of red giant phase and/or thermal pulses). This behaviour is consistently observed across all three levels of explored model complexity (i.e., SimA, SimB, and SimC).

Although the WRLO phase is predicted to yield high mass accretion efficiencies, of a few times $\eta \sim 0.1$, our modelled systems are compact enough to achieve comparable efficiencies under a standard wind accretion regime. This is illustrated by the models shown in Fig.~\ref{fig:a_eta} (see also Fig.~\ref{fig:eta_wrlo}), which exhibit values of $\eta \lesssim 0.4$ following the RGB phase, despite operating under standard wind accretion. Therefore, contrary to previous suggestions, standard wind accretion can produce similar mass accretion efficiencies as the WRLO mechanism, depending on the system's configuration and the evolutionary behaviour of the donor star’s wind parameters.

In addition to the close orbital configuration (i.e., high $\varv_\mathrm{o}$), the primary factor responsible for the high mass accretion efficiencies observed in our wind accretion simulations is the slow wind velocity ($\varv_\mathrm{w} \lesssim$ 2–12 km s$^{-1}$) predicted by our stellar evolution models (see Appendix~\ref{app:vwind_mdot} and Paper~I). This condition ensures that our systems evolve in the $\varw = \varv_\mathrm{w}/\varv_\mathrm{o} < 1$ regime (see Fig.~\ref{fig:eta_w}), naturally leading to high values of the accretion efficiency $\eta$ as predicted by Eq.~(\ref{eq:TejedaToala}). In contrast, it seems that \citet{Vathachira2025} adopted a constant wind velocity of 20 km s$^{-1}$, placing their models in the $\varw > 1$ regime, where wind accretion efficiencies are significantly lower, as illustrated by the dashed line in Fig.~\ref{fig:eta_w}.

We find that only symbiotic binaries with high-mass WD ($m_{2,0} > 1$ M$_\odot$) and relatively massive donor stars ($m_{1,0} = 2$--3 M$_\odot$) are able to reach the Chandrasekhar limit when WRLO is included. Only 11 per cent of the full simulation set achieves this threshold, all of these successful cases rely on WRLO to accumulate sufficient mass. This highlights the importance of incorporating WRLO in models to accurately evaluate the potential of such systems as Type Ia supernova progenitors, in agreement with the conclusions of \citet{Il2019}. Additionally, we predict that simulations halted at the onset of RLO may represent another population of systems capable of reaching the Chandrasekhar mass limit \citep[for example, the case of T CrB;][]{Munari2025}. Most of these cases are compact binaries ($a_0 = 2-3$ AU) evolving through the TPAGB phase, regardless of the initial masses of the donor and accretor. While we do not model the RLO phase in this work, since it requires a fully hydrodynamical treatment, which we plan to explore in future studies, these systems may eventually reach the critical mass during the RLO phase, potentially triggering additional evolutionary pathways such as common envelope channel \citep[][]{Iben1993,Taam2000,Ivanova2013}.

\begin{figure}
\includegraphics[width=\linewidth]{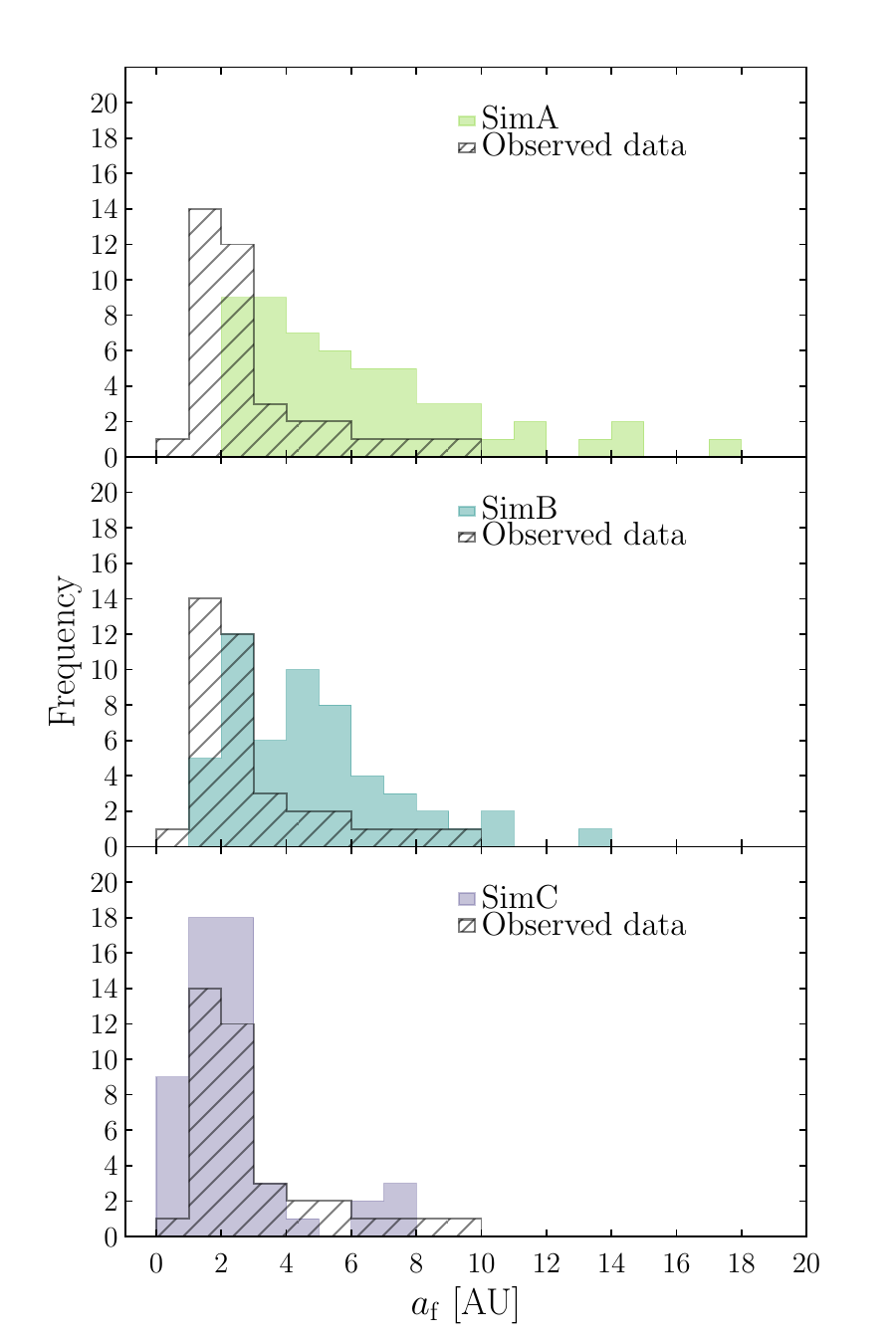}
\caption{Distribution of final semimajor axes ($a_\mathrm{f}$) from all the 162 simulations. Hatched bars indicate the semimajor axes of observed symbiotic systems with $a\leq$ 10 AU as listed in the {\it New Online Database of Symbiotic Variables} \citep{Merc2019} (see Table \ref{tab:objects}). Panels from top to bottom correspond to SimA, SimB, and SimC.}
\label{fig:af_hist2}
\end{figure}

Simulations incorporating different levels of physical complexity yield distinct orbital configurations. In general, models that include more physical processes result in more compact systems. To illustrate this trend, Fig.~\ref{fig:af_hist2} shows the final semimajor axis ($a_\mathrm{f}$) distributions from our simulations, grouped by increasing physical complexity. The top panel, corresponding to simulations with only wind accretion (SimA), exhibits a broad range of $a_\mathrm{f}$ values, from 1 to 18 AU. A significant fraction of SimA models, approximately 33 per cent, evolve to orbital separations greater than 7 AU. When wind drag is included (SimB), this fraction drops to 17 per cent (Fig.\ref{fig:af_hist2}, middle panel). In contrast, simulations that also include tidal interactions (SimC) result in even more compact systems, with only 6 per cent having final orbits exceeding 7 AU (Fig.\ref{fig:af_hist2}, bottom panel).

For comparison, the hatched bars in Fig.~\ref{fig:af_hist2} show the semimajor axis distribution of known symbiotic systems with $a <$ 10 AU, taken from the {\it New Online Database of Symbiotic Variables}\footnote{\url{https://sirrah.troja.mff.cuni.cz/~merc/nodsv/}} \citep{Merc2019}. Additional properties of the observed systems are provided in Table~\ref{tab:objects} in Appendix~\ref{app:observations}.
Most observed systems are compact, with $a$ between 1 and 3 AU.

Fig.~\ref{fig:af_hist2} shows that the final orbital configuration distribution created by SimC simulations agree best with observed systems, suggesting that observed compact symbiotic systems with $a < 4$ AU may be undergoing orbital decay driven by tidal forces. However, this result should be treated with caution, as the direct comparison between our numerical results and observations relies on the assumption that all initial conditions adopted in the simulations are equally probable. In addition, we note that stellar rotation is not included in the current simulations, yet it is expected to play a significant role in the dynamical evolution of such systems. Tidal interactions may lead to synchronization (tidal locking), which could reduce or even halt the inward migration of the companion \citep[e.g.][]{Fleming2019}. Furthermore, rotation could alter the geometry of the donor star's mass loss, promoting wind focusing toward the orbital plane and thereby enhancing mass-transfer efficiency \citep{Skopal2015}. 
Overall, these results suggest that tidal forces need to be considered as a key physical component in the study of compact symbiotic systems.

 We note that the orbital evolution of interacting binaries can also be, in principle, studied solely using orbit-averaged formalism and the binary-evolution capabilities of {\sc mesa}. The orbit-averaged formalism provides an efficient way to compute long-term changes in the orbital elements \citep[e.g.][]{Sepinsky2009}, but it cannot capture the short-scale system's response during episodes of rapid stellar evolution, particularly the variation in mass-loss rate and radius expansion predicted by the detailed stellar evolution models during thermal pulses. Orbit-averaged calculations are expected to naturally smooth the details in the calculations. 

In contrast, direct integration yields detailed orbital elements and accretion histories at physically meaningful stopping points, such as RLO or when the accretor approaches the Chandrasekhar limit, facilitating future hydrodynamical studies. Moreover, it allows us to identify specific times of transitions between distinct accretion regimes, including wind accretion and WRLO, that depend sensitively on the instantaneous orbital configuration.

The binary module of {\sc mesa} offers a complementary tool for studying interacting binaries through orbit-averaged equations coupled with detailed stellar-structure evolution \citep[e.g.][]{Paxton2015,Henneco2024}. Within that framework, different physical processes, such as tides, mass transfer, and angular momentum loss, can be selectively included, in close analogy to our treatment. Recent efforts have also demonstrated the consistency and compatibility of how stellar evolution in binary module and direct N-body integrations can be effectively coupled \citep[e.g.][]{Xiang2025}. This further supports the view that {\sc mesa} and {\sc rebound}/{\sc reboundx} offer complementary strengths.

While the present study focuses on symbiotic binaries in circular orbits, this represents a useful first approximation, as tidal forces are expected to act efficiently toward circularization over time \citep[e.g.][]{Zahn1977,Dewberry2025}. Nevertheless, some observed systems do exhibit small to moderate eccentricities (see Table~\ref{tab:objects}), which may influence their long-term evolution. Interestingly, \citet{Marinovic2008} calculations show that tidally enhanced mass loss during the AGB phase can excite or sustain orbital eccentricity, partially counteracting tidal damping. Modelling eccentric orbits could provide further insight into how tidal interactions, combined with stellar rotation and mass transfer, might prolong the WRLO phase or enhance the likelihood of reaching the Chandrasekhar limit. Exploring the role of eccentricity in this context is a natural extension of the current work that will be addressed in future studies.

 Finally, we note that in the present study we assume that the evolving red giant donor star and the WD companion evolve largely independently, without significantly affecting one another. Although the modelled systems are relatively compact, the accretion rate at the start of the simulations is negligible, supporting this approximation.

\section{Summary}
\label{sec:summary}

We studied the impact of wind accretion in evolving compact (2 AU $\leq a_0 \leq 7$ AU) symbiotic systems. We modelled the evolution of the mass losing star with {\sc mesa} and its dynamical interactions with an accreting companion was followed through $N$-body simulations with the code {\sc rebound}. The initial masses of the accreting WDs were set to $m_{2,0}=0.7$, 1.0, and 1.2 M$_\odot$, which orbited evolving Solar-like stars with masses of $m_{1,0}=1$, 2, and 3 M$_\odot$. 

The compact symbiotic binaries were modelled under various physical configurations, incorporating wind accretion (SimA), wind drag (SimB), and tidal interactions (SimC). The wind accretion is modelled adopting the modified BHL prescription recently presented by \citet{TejedaToala2025}, however, the evolving systems are allowed to experience a WRLO accretion regime with a numerical recipe from \citet{Abate2013} when $R_\mathrm{cond} \geq R_\mathrm{Roche}$.

Our main findings can be summarized as follow:
\begin{itemize}
    
\item Of the 162 modelled systems, 11 per cent (18/162) push the mass of the accretor to the Chandrasekhar limit and about 29 percent (46/162) evolve through the end of the TPAGB phase without reaching this limit. The rest of them, 60 percent (98/162), are stopped at the beginning of the RLO phase.

\item The RLO phase is not modelled here because such complex accretion phase would require hydrodynamical simulations to study those systems. It is very likely that such channel will be also efficient enough to take the accreting WD to the Chandrasekhar limit, particularly because the stopping condition took place during the period with the highest mass-loss, the TPAGB phase.

\item Our results support previous suggestions that a single evolving symbiotic system can transition between the standard wind accretion and WRLO regimes. The WRLO is active during periods of high mass loss from the donor star, the peak of red giant phase and/or thermal pulses, but it does not have a long duration.

\item Depending on several factors (e.g., the masses of the two stars, their initial orbital configuration, and the evolution of the stellar wind properties of the mass losing star), we found that the standard wind accretion mechanism is also able to produce high mass accretion efficiencies (of a few times of 0.1) as found for the WRLO phase. In contrast, phases of standard wind accretion have longer duration.

\item A comparison with observations of compact symbiotic systems suggests that they should be undergoing orbital decay driven by tidal forces. If this is to be the case, tidal forces need to be considered when studying compact symbiotic systems in order to make more realistic estimations and better prescriptions of their evolutionary pathways.

\end{itemize}

\section*{Acknowledgements} 

We thank the anonymous referee for taking the time reviewing our work and for comments and suggestions that improved our original manuscript. R.F.M. thanks UNAM DGAPA (Mexico) and SECIHTI (Mexico) for postdoc fellowships. J.A.T. and J.B.R.-G. acknowledge support from the UNAM PAPIIT project IN102324. J.A.T. thanks the staff of Facultad de Ciencias de la Tierra y el Espacio of Universidad Aut\'{o}noma de Sinaloa (FACITE-UAS, Mexico) for their support during a research visit. This work has made extensive use of NASA's Astrophysics Data System.



%\bibliographystyle{mnras}

\section*{DATA AVAILABILITY}
The model results underlying this article will be shared on reasonable request to the corresponding author.

\appendix

\section{Mass-loss rate and stellar wind velocity evolution}
\label{app:vwind_mdot}

\begin{figure*}
\begin{center}
\includegraphics[width=\linewidth]{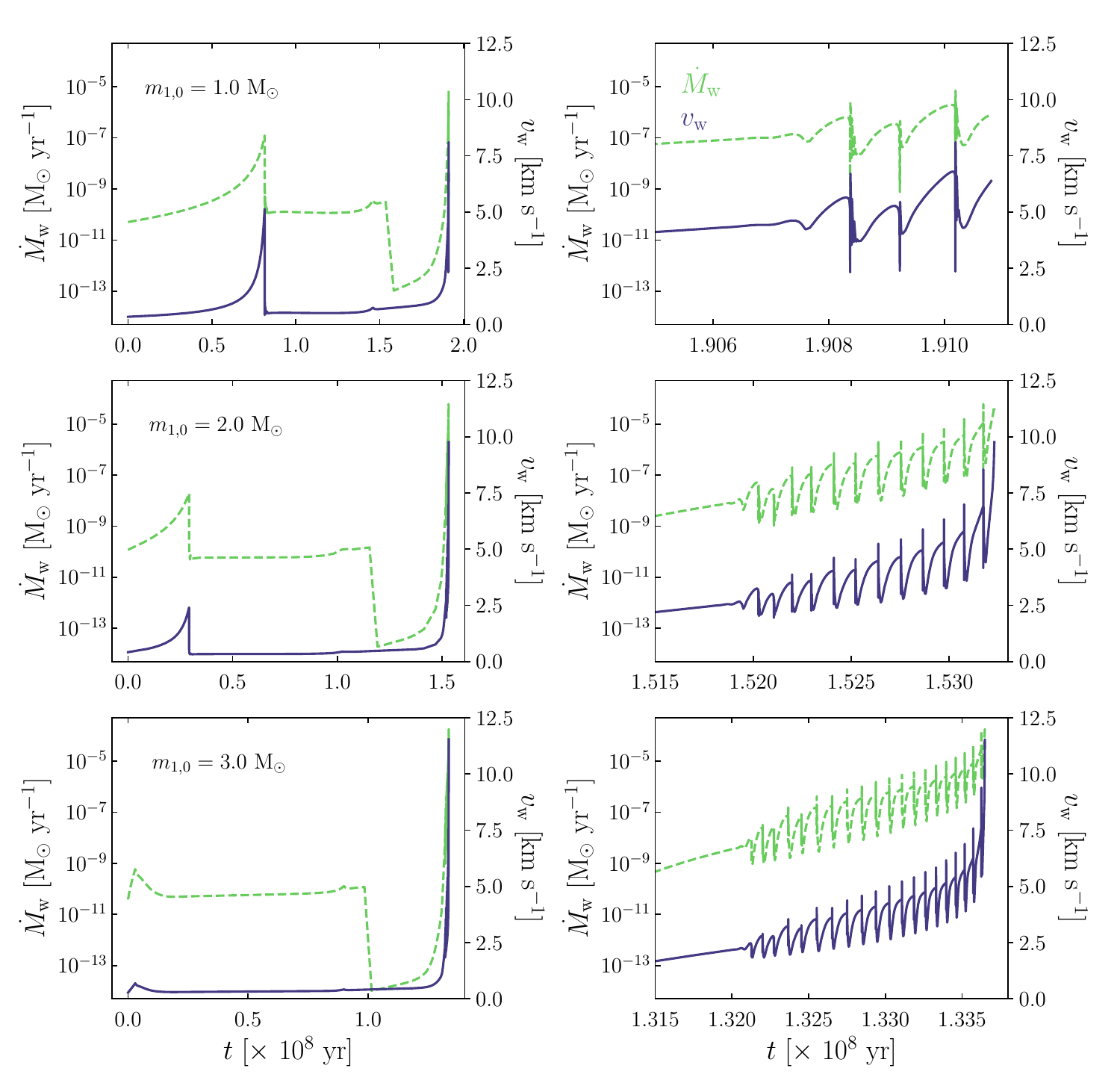}\\
\end{center}
\caption{Terminal wind velocity and mass-loss rate as a function of time for the three stellar evolution models with initial masses of 1.0 (top), 2.0 (middle), and 3.0 M$_\odot$ (bottom). The left panels show the full evolutionary sequence, from the RGB to the end of the TPAGB phases, while the right column panels provide a zoom-in on the TPAGB phase. In all panels, the dashed and solid lines show the evolution of $\dot{M}_\mathrm{w}$ and $\varv_\mathrm{w}$, respectively.}
\label{fig:vel_wind_mdot}
\end{figure*}

Fig.~\ref{fig:vel_wind_mdot} presents the evolution of the wind velocity ($\varv_\mathrm{w}$, solid lines) and mass-loss rate ($\dot{M}_\mathrm{w}$, dashed lines) used in our calculations. The left panels of the figure show the full evolution sequence, from the onset of the RGB to the end of the TPAGB phase, while the right panels zoom-in on the TPAGB phase alone. From top to bottom, the panels correspond to stellar models with initial masses of $m_{1,0} = 1.0$, $2.0$, and $3.0~\mathrm{M}_\odot$, respectively. Further details of the evolution of these parameters are given in Paper~I.

\section{Entering the WRLO phase}
\label{app:WRLO}

In this appendix we use the simulation examples presented in Fig.~\ref{fig:a_eta} and \ref{fig:a_eta2} of Section~\ref{sec:results} to highlight the times they switch from standard wind accretion to WRLO phase. 

\begin{figure*}
\begin{center}
\includegraphics[width=0.95\linewidth]{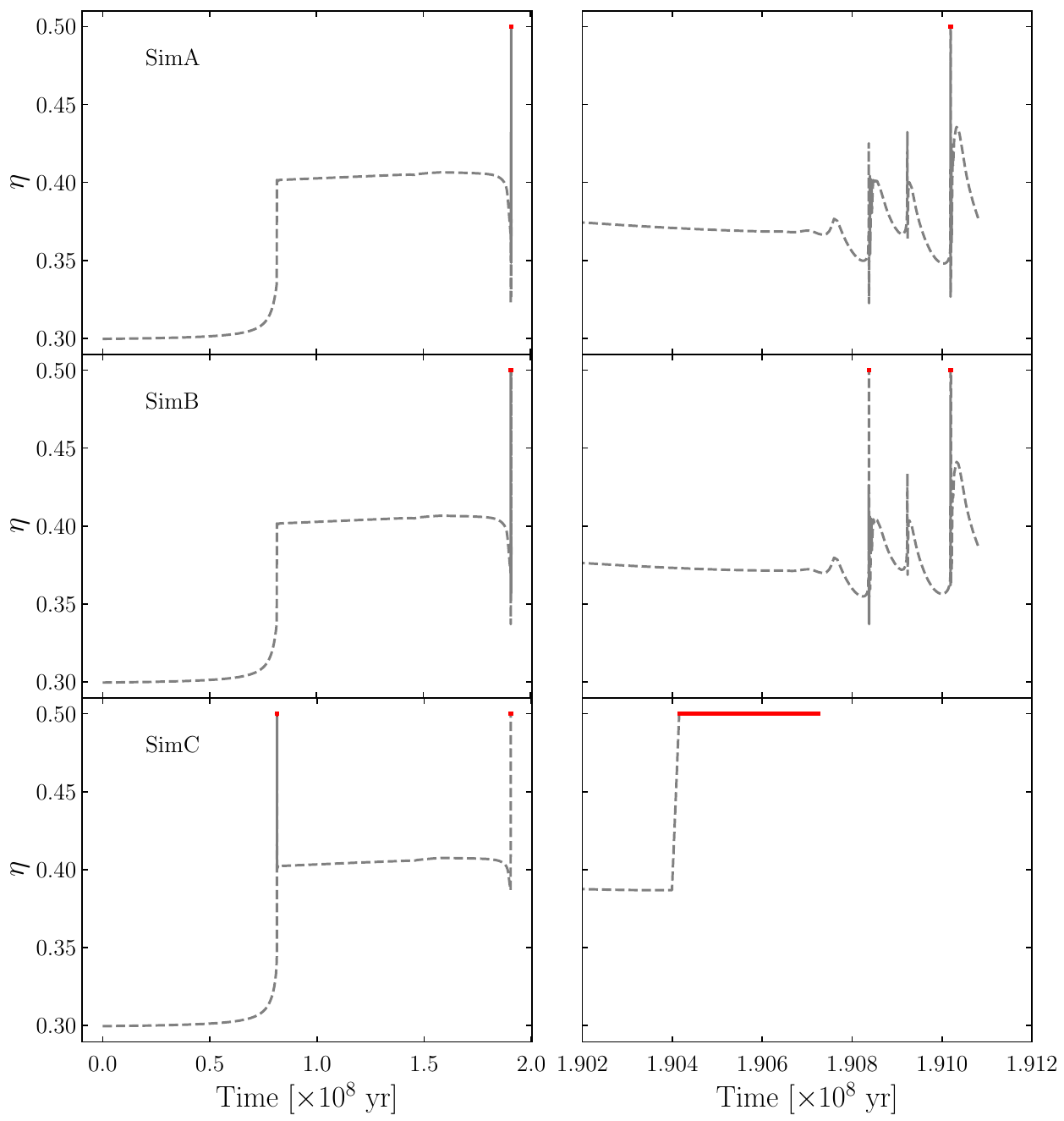}\\
\end{center}
\caption{Evolution of the wind accretion efficiency $\eta$ for SimA (top), SimB (middle), and SimC (bottom) of a symbiotic binary system with $m_{1,0} = 1.0$~M$_\odot$, $m_{2,0} = 1.2$~M$_\odot$, and $a_0=6.0$ AU. The left panels display the full integration, while the right panels focus on the TPAGB phase (see also Fig.~\ref{fig:a_eta}). The (gray) dashed lines mark the time intervals when the standard wind accretion efficiency is applied while the (red) solid lines indicate the periods when the WRLO phase is activated.}
\label{fig:eta_wrlo}
\end{figure*}

Fig.~\ref{fig:eta_wrlo} shows the results from models with $m_{1,0}=1.0$ M$_\odot$, $m_{2,0}= 1.2$ M$_\odot$, and $a_0 = 6.0$ AU. The (red) solid line illustrates times when the WRLO phase is active, when $R_\mathrm{cond} \geq R_\mathrm{Roche}$. The left panels correspond to the total evolution of the systems (SimA - top, SimB - middle, and SimC - bottom) whilst the right panels zoom into the TPAGB phase.

\begin{figure*}
\begin{center}
\includegraphics[width=0.95\linewidth]{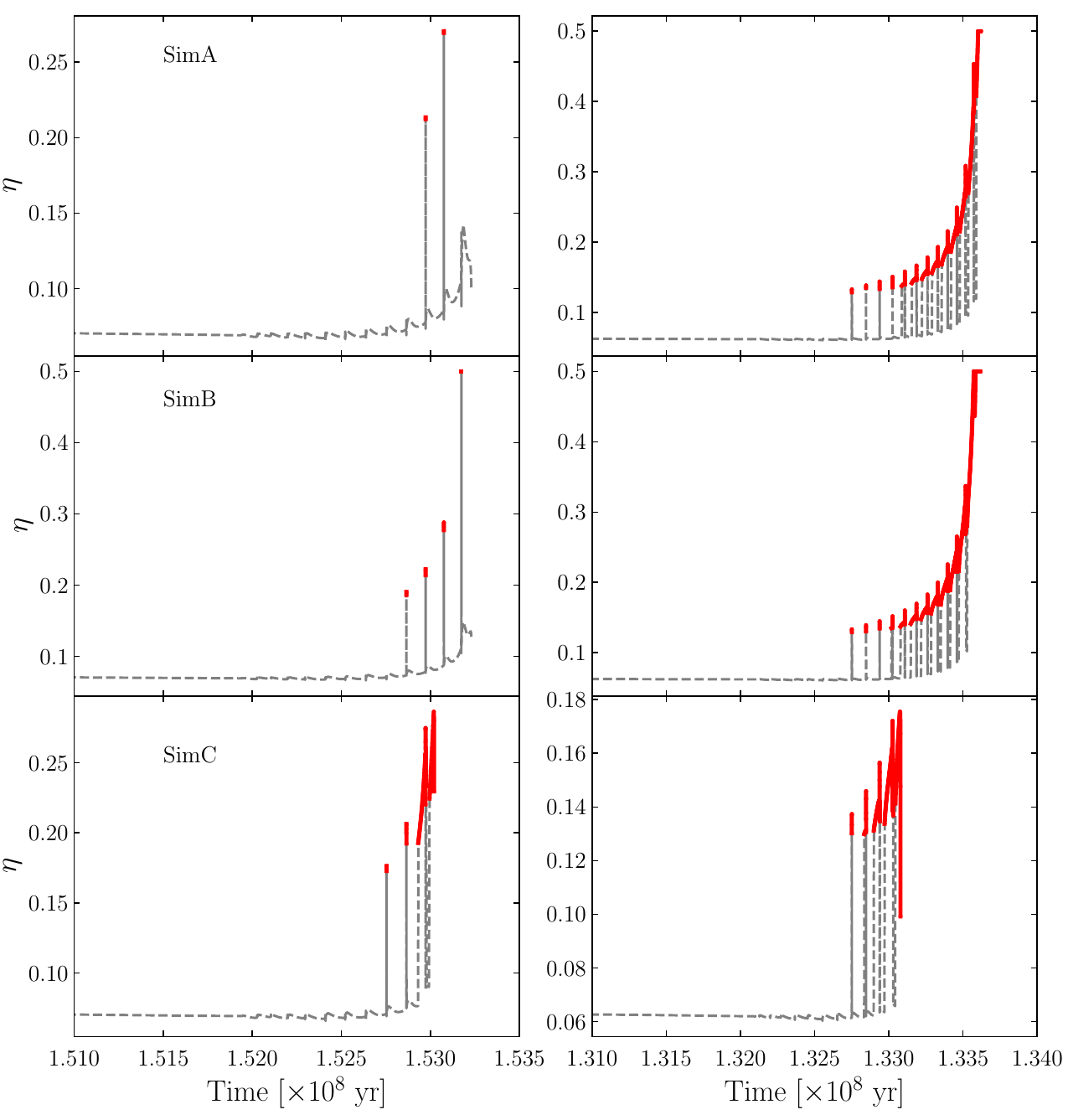}\\
\end{center}
\caption{Same as Fig.~\ref{fig:eta_wrlo} but for simulations with $m_{1,0} = 2.0$~M$_\odot$ and $m_{2,0} = 0.7$~M$_\odot$ (left panels) and  $m_{1,0} = 3.0$~M$_\odot$ and $m_{2,0} = 1.0$~M$_\odot$ (right panels). The top, middle and bottom rows correspond to SimA, SimB, and SimC, respectively. For simplicity we only show their evolution during the TPAGB phase. See also Fig.~\ref{fig:a_eta2} for more details.}
\label{fig:eta_wrlo2}
\end{figure*}

Fig.~\ref{fig:eta_wrlo2} shows the results from simulations with evolving mass losing stars with initial masses of $m_{1,0}=2.0$ and 3.0 M$_\odot$ with companions with initial masses of $m_{2,0} = 0.7$ and 1.0 M$_\odot$, respectively, orbiting at $a_0=$ 6 AU.

\section{Observed symbiotic systems}
\label{app:observations}

Table~\ref{tab:objects} lists a sample of symbiotic systems from the {\it New Online Database of Symbiotic Variables}\footnote{\url{https://sirrah.troja.mff.cuni.cz/~merc/nodsv/}} \citep{Merc2019}. We selected the best characterised compact systems. That is, systems with estimated masses ($m_1$ and $m_2$) and orbital period ($T$) that have estimated semi-major axis ($a$) smaller than 10 AU. These symbiotic systems are used to create the histogram of observed data of Fig.~\ref{fig:af_hist2}.

\begin{table}
\begin{center}
\footnotesize
\caption{Physical and orbital parameters estimated for symbiotic systems with semimajor axis $a<10$ AU as listed in the {\it New Online Database of Symbiotic Variables} \citep{Merc2019}. The columns list the name of the binary, mass of the primary (donor) $m_1$, mass of the secondary (accretor) $m_2$, orbital period $T$, orbital separation $a$ and eccentricity $e$, from left to right, respectively.} 
\begin{tabular}{lccccc}
\hline
Object & $m_1$      & $m_2$       & $T$  & $a$ & $e$\\
 & [M$_\odot$]& [M$_\odot$] & [yr] &  (AU) & \\
\hline
    AE Ara &              2.00 &            0.51 &     2.00 &      2.2    & 0 \\
    AE Cir &              1.10 &           1.00 &     0.94 &      1.2  & \dots\\
    AG Dra &              1.20 &          0.50 &     1.51 &      1.6  & 0.0006\\
    AG Peg &              2.60 &            0.65 &     2.23 &      2.5 & 0.11 \\
    AR Pav &              2.00 &            0.75 &     1.65 &      2.0 & 0\\
    AX Per &              3.00 &            0.60 &     1.86 &      2.3 & 0\\
    BF Cyg &              1.50 &            0.40 &     2.07 &      2.0 & 0 \\
     BX Mon &              3.70 &           0.55 &     3.78 &      3.9 & 0.444\\
    CH Cyg &              2.20 &            0.56 &    15.58 &      8.8 & 0.122\\
    CI Cyg &              2.40 &           0.50 &     2.34 &      2.5 & 0.109 \\
    CQ Dra &              5.00 &            0.85 &     4.66 &      5.1 & 0.3\\
    EG And &              1.46 &            0.40 &     1.32 &      1.5 & 0 \\
    ER Del &              3.00 &            0.70 &     5.72 &      5.0 & 0.228\\
    FG Ser &              1.70 &            0.60 &     1.78 &      1.9 & 0\\
    FN Sgr &              1.50 &            0.70 &     1.55 &      1.7 & 0 \\
    HD 330036 &              4.46 &         0.54 &     4.59 &      4.8 & \dots\\
    Hen 3-461 &              1.50 &           0.88 &     6.22 &      4.5 & 0.4\\
    Hen 3-828 &           1.5 &             0.6 &       1.81 &      1.9 & 0 \\
    IV Vir &              0.90 &            0.42 &     0.77 &      0.9 & 0 \\
    KX TrA &              1.00 &            0.41 &     3.31 &      2.5 & 0.29\\
    LT Del &              1.00 &            0.57 &     1.24 &      1.4 & 0.4\\
    PU Vul &              1.00 &            0.50 &    13.42 &      6.5 & 0.16\\
    RS Oph &              0.68$-$0.8 &      1.2$-$1.4&  1.24 &      1.4$-$1.5& 0\\  
    RW Hya &              1.60 &            0.48 &     1.01 &      1.3 & 0 \\
    St 2-22 &              2.80 &            0.80 &     2.51 &      2.9 & 0.16\\
    SY Mus &              1.30 &            0.43 &     1.71 &      1.7 & 0 \\
     T CrB &              2.10 &            1.30 &     0.62 &      1.0 & 0\\
     TX CVn &              3.50 &            0.40 &     0.55 &      1.1 & 0.6 \\
     V443 Her &              2.50 &            0.42 &     1.64 &      2.0 & 0\\
    V455 Sco &              1.10 &            0.60 &     3.85 &      3.0 & 0\\
    V471 Per &              2.30 &          0.80 &    17.00 &      9.4 & \dots\\
     V694 Mon &              1.00 &          0.90 &     5.28 &      3.8 & 0.68$-$0.82\\
    V745 Sco &            1.00 &            1.38 &     1.40 &      1.7 & \dots\\
    V934 Her &              1.60 &          1.35 &    12.02 &      7.6 & 0.354\\
    V1261 Ori &              1.65 &         0.55 &     1.75 &      1.9 & 0.07 \\
    V1329 Cyg &              2.10 &            0.70 &     2.62 &      2.7 & 0 \\
    V3890 Sgr &            1.05 &          1.35 &     2.05 &      2.2 & \dots \\
     Z And &              2.00 &          0.65 &     2.08 &      2.3 & 0 \\
\hline
\end{tabular}
\label{tab:objects}
\end{center}
\end{table}

\end{document}